\documentclass{evn2004}
\usepackage{graphicx}
\usepackage{txfonts}
\begin{document}
\setcounter{page}{295}

   \title{A 32 m Parabolic Antenna in Peru At 3,370m of Altitude}
   \author{Jos\'e Ishitsuka\inst{1}\thanks{E-mail: pepe@hotaka.mtk.nao.ac.jp}
           \and Mutsumi Ishitsuka\inst{2}
           \and Norio Kaifu\inst{1} 
           \and Shoken Miyama\inst{1} 
           \and Makoto Inoue\inst{1} 
           \and Masato Tsuboi\inst{1,3} 
           \and Masatoshi Ohishi\inst{1} 
           \and Kenta Fujisawa\inst{5} 
           \and Takashi Kasuga\inst{6} 
           \and Keisuke Miyazawa\inst{1} \thanks{Retired}
           \and Shinji Horiuchi\inst{4}
          }
\institute{National Astronomical Observatory, Japan
           \and Instituto Geof\'{\i}sico del Per\'u (IGP), Per\'u
           \and Nobeyama Radio Observatory, Japan
           \and Square Kilometer Array (SKA), Australia
           \and Yamaguchi University, Japan, 6 Hosei University, Japan
          }

\abstract{
At the altitude of 3,370 m on the Peruvian Andes, a 32m antenna owned
by the telecommunications company Telef\'onica del Per\'u will be
transformed to a Radio Telescope, it would be transferred to the
Geophysical Institute of Peru (IGP). The parabolic antenna was
constructed in 1984 by Nippon Electric Co. (NEC) and worked as an
INTELSAT station until 2000. A team of the National Observatory of
Japan (NAOJ) evaluated the antenna in 2003 and reported its
availability to be used as a Radio Telescope. In collaboration of the
NAOJ a 6.7 GHz receiver is under construction and will be installed
within this year. Initially the telescope as a single dish will monitor
and survey Methanol Maser of YSO, higher frequencies equipment and VLBI
instruments will be considered. The antenna will be managed by the IGP
and used by universities in Peru, becoming a VLBI station will be a
grate contribution to astronomy and geodetic community.
        }

\maketitle

%
%

\section{The Antenna}

The INTELSAT station was built between 1984 and 1985 by the ENTEL-PERU
(Empresa Nacional de Telecumunicaciones - Per\'u) the national
communications enterprise of Peru, the antenna system was built by the
Nippon Electric Co., and then in 1993 the national telecommunications
enterprise was transferred to the private company Telef\'onica del
Per\'u. In 2000 the antenna station stopped operations as an INTELSAT
station.

Actually the antenna station belongs to this private company and
negotiations are in progress to transfer the antenna station to the
IGP.

\section{Location}

The Sicaya antenna station is located on a small hill in a beautiful
open flat valley that is similar to Owens Valley in California or
Nobeyama in Japan where world famous radio observatories are located.
The antenna looks still in very good condition without no apparent rust
probably due to its location at 3,370 m of high altitude and being far
away from sea side.  The structure of the antenna is well suited to
upgrade receivers in future.\\

\noindent
Latitude: -12$^{\circ}$02'15", \\
longitude: -75$^{\circ}$17'39",\\ 
altitude: 3,370 m.\\

\noindent
Weather: \\
Annual precipitation: 700 mm in average,\\ 
average temperature: 12$^{\circ}$C,\\
average humidity: 61 \%, no snow.

\begin{figure*}
  \centering 
  \includegraphics[angle=0,width=\textwidth]{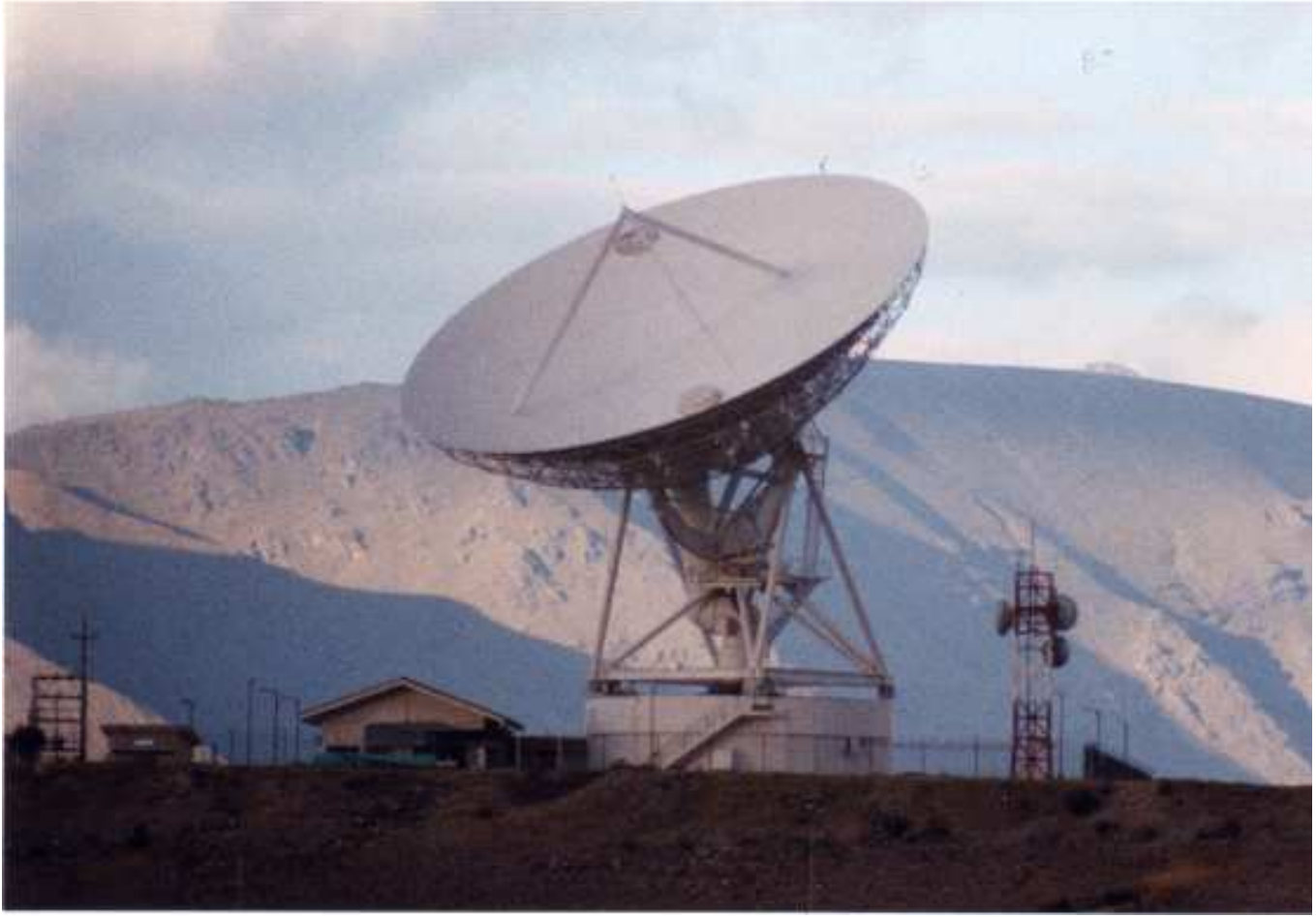}
  \caption{The 32m Sicaya antenna is located at the altitude of 3,370 m
  on the Peruvian Andes.}
  \label{fig:Sicaya}
\end{figure*}

\section {First Step}

As the first step the antenna will be equipped with a 6.7 GHz
receiver. The 6.7 GHz receiver is under construction and planed to be
ready for this spring.

Initially we are planning to monitor and survey Methanol masers on
Young Stellar Objects. Methanol masers are good tracers around the
young objects that can be used to disentangle unknown physical
conditions of stellar surroundings.

\section {VLBI}

Once the antenna is equipped and allowed to observe at higher
frequencies, it will became an important VLBI station in South America,
due to the unique location and altitude will contribute enormously to
for instance the VLBA array (Horiuchi et al. 2003). In the future it
will become a promising geodetic station.

\section {Conclusion}

The 32 m antenna in Peru will be transformed into a radio telescope. As
the first step, as a single dish telescope the antenna will be equipped
with a 6.7 GHz receiver to monitor and survey methanol masers of Young
Stellar Objects. The antenna will become a powerful mean to develop and
settle radio astronomy in Peruvian institutions and universities. In
the future, receiving higher frequencies and participating as a VLBI
station are in mind.

\end{document}